\documentclass[%
reprint,
superscriptaddress,
preprintnumbers,
amsmath,amssymb,aps, 
prb,
]{revtex4-2}

\usepackage[utf8]{inputenc}
\usepackage[version=3]{mhchem}
\usepackage{graphicx}
\usepackage{dcolumn}
\usepackage{bm}
\usepackage{float}
\usepackage[usenames,dvipsnames]{color}

\begin{document}


\title{Direct evidence of ferromagnetism in \ce{MnSb2Te4}}

\author{Wenbo Ge}
\affiliation{Department of Physics and Astronomy, Rutgers University, Piscataway, New Jersey 08854, USA}
\author{Paul M. Sass}
\affiliation{Department of Physics and Astronomy, Rutgers University, Piscataway, New Jersey 08854, USA}
\affiliation{Pacific Northwest National Laboratory, Richland, Washington, 99352, USA}
\author{Jiaqiang Yan}
\affiliation{Materials Science and Technology Division, Oak Ridge National Laboratory, Oak Ridge, Tennessee 37831, USA}
\affiliation{Department of Materials Science and Engineering, University of Tennessee, Knoxville, Tennessee 37996, USA}
\author{Seng Huat Lee}
\affiliation{2D Crystal Consortium, Materials Research Institute, Pennsylvania State University, University Park, PA 16802, USA}
\affiliation{Department of Physics,  Pennsylvania State University, University Park, PA 16802, USA}
\author{Zhiqiang Mao}
\affiliation{2D Crystal Consortium, Materials Research Institute, Pennsylvania State University, University Park, PA 16802, USA}
\affiliation{Department of Physics,  Pennsylvania State University, University Park, PA 16802, USA}
\author{Weida Wu}
\email{wdwu@physics.rutgers.edu}
\affiliation{Department of Physics and Astronomy, Rutgers University, Piscataway, New Jersey 08854, USA}%

\begin{abstract}
We report the magnetic imaging of ferromagnetic domains in the van der Waals single crystal \ce{MnSb2Te4} from two different sources using cryogenic magnetic force microscopy. The magnetic field dependence of the domains reveals very weak pinning of domain walls in \ce{MnSb2Te4}, resulting in a negligibly small magnetic hysteresis loop. The temperature dependence of the domain contrast reveals a mean field like behavior, in good agreement with that of bulk magnetization measurements. 
\end{abstract}

\maketitle


\section{\label{sec:level1}Introduction}
The interplay between magnetism and topological electronic states has led to many fascinating  quantum phenomena, such as the quantum anomalous Hall (QAH) effect and axion insulator state. The QAH state was realized for the first time in the magnetically doped topological insulator \ce{(Bi\textrm{,} Sb)2Te3} \cite{Chang2013Science}. The introduction of magnetic dopants breaks local time reversal symmetry and opens an exchange gap at the Dirac point of the topological surface band, leading to a dissipationless chiral edge mode with no external magnetic field \cite{Tokura2019NRP,Qi2008PRB,Yu2010Science,Chang2013Science,Chen2010Science}. However, the magnetic inhomogeneity inherent to doping limits the full quantization to ultra-low temperatures \cite{Lachman2015SciAdv,Grauer2015PRB,Lee2015PNAS,Chang2014PRL}. The recently synthesized \ce{MnBi2Te4}-family compounds are van der Waals materials that host intrinsic magnetism and topologically nontrivial band structure \cite{Yan2019PRM,Otrokov2019Nature,Gong2019CPL,Li2019SciAdv,Otrokov20172DMaterials,Zeugner2019ChemistryofMaterials,Lee2019PhysRevResearch,Cui2019PRB}. The magnetism of this compound comes from the long-range order of Mn layers instead of magnetic doping. As an intrinsic magnetic TI, \ce{MnBi2Te4} provides a clean platform to realize the QAH effect at elevated temperatures and to explore other exotic topological phases.  The structure of the \ce{MnBi2Te4} septuple-layer can be viewed as a Mn-Te bilayer being inserted into the middle of a quintuple layer of \ce{Bi2Te3}. Each Mn$^{2+}$ ion carries 5~$\mu_\textrm{B}$ magnetic moment in the localized magnetism approximation. The intra-plane exchange coupling is ferromagnetic while the inter-plane coupling is antiferromagnetic, which results in an A-type antiferromagnetic order with uniaxial anisotropy \cite{Otrokov2019Nature,Yan2019PRM}. First principle calculations predict that an odd number of septuple-layers can realize the QAH effect and an even number can realize an axion insulator state \cite{Otrokov2019PRL}. Both phenomena were recently reported in odd and even layers of exfoliated flakes of \ce{MnBi2Te4} single crystals above 1~K \cite{Deng2020Science, Liu2020Nature}. However, the topological nature of the electronic phases of even and odd layers remains controversial \cite{ovchinnikov2020arxiv}.

\ce{MnSb2Te4} is isostructural to \ce{MnBi2Te4} \cite{Yan2019prb}. Its ideal magnetic ground state is also A-type antiferromagnetic \cite{Gong2019CPL}. However, a recent study revealed that the magnetism in \ce{MnSb2Te4} can be tuned from antiferromagnetic to ferromagnetic by changing the Mn/Sb site mixing \cite{liu2020arxiv}. This study pointed out a potential route to tune the magnetism in \ce{MnBi2Te4} to realize a robust QAH effect at high temperature. In ferromagnetic \ce{MnSb2Te4}, the Mn atoms mixed into the Sb sites order antiparallel to neighboring Mn layers and mediate a ferromagnetic inter-septuple-layer coupling. Density functional theory (DFT) calculations show that ferromagetic \ce{MnSb2Te4} may host Weyl points despite the fact that synthesized ferromagnetic \ce{MnSb2Te4} requires significant site mixing leading to a topologically trivial band structure \cite{liu2020arxiv}. Engineering the magnetism and maintaining the band topology of \ce{MnSb2Te4} and \ce{MnBi2Te4} requires further study of their electronic band structure and magnetic properties. Neutron diffraction, transport and magnetization studies have been reported on ferromagnetic \ce{MnSb2Te4}, yet, the direct visualization of the ferromagnetic state in \ce{MnSb2Te4} is still lacking \cite{liu2020arxiv,Wimmer2020arxiv}. 

Here, we report cryogenic magnetic force microscopy (MFM) of ferromagnetic \ce{MnSb2Te4} single crystal samples from two different sources. Micron size ferromagnetic domains are visualized on both samples after zero-field cooling (ZFC) below $T_\textrm{C}\approx 33$~K. We observed typical soft ferromagnetic domain behavior with very weak domain wall pinning. The magnetic field dependence of domain population shows little hysteresis, in good agreement with that of bulk magnetization. The temperature dependence of domain contrast follows mean field behavior, in good agreement with that of the saturation magnetization. 

\begin{figure*}[hbtp]
    \centering
    \includegraphics[width=0.9\textwidth]{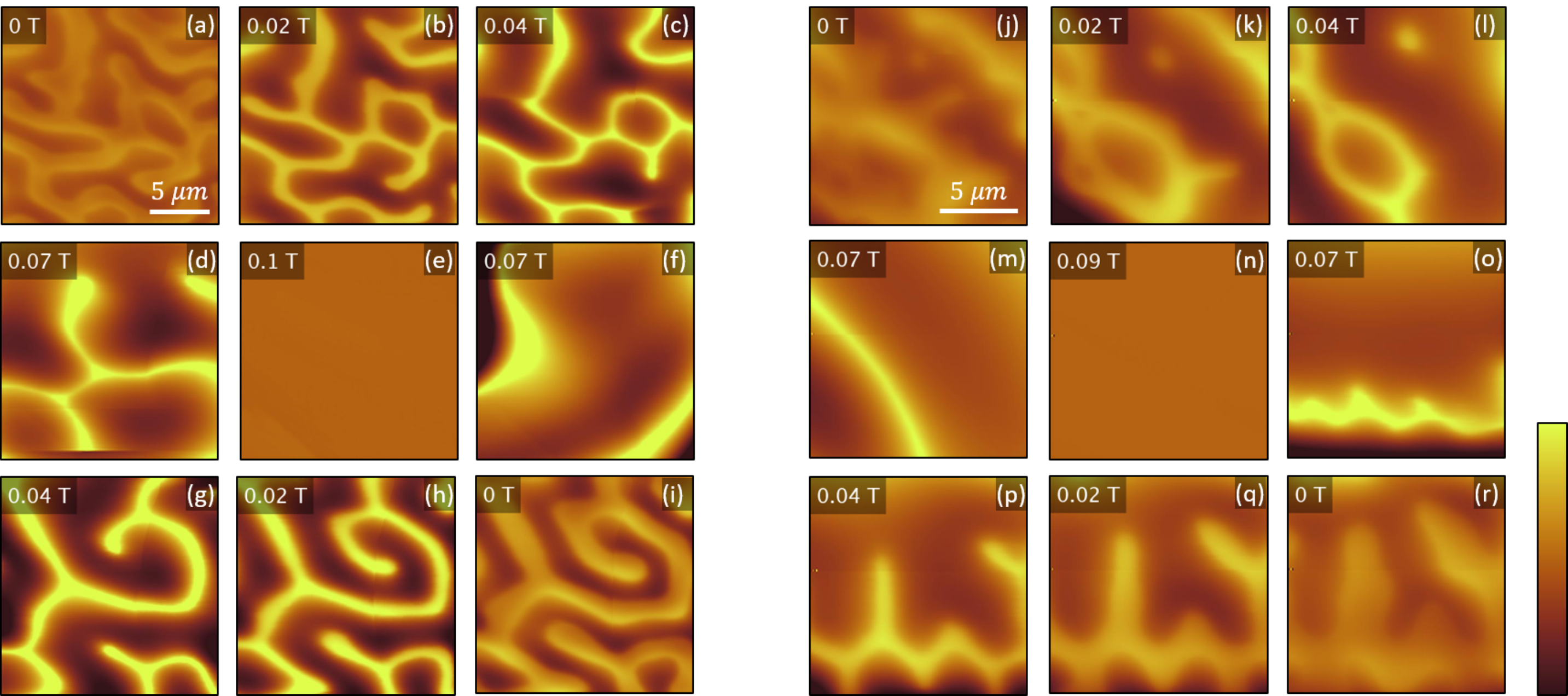}
    \caption{(Color Online) Field dependent MFM images of two samples synthesized by different groups. (a)-(i) Sequential MFM images of S1 (Penn State) measured at 6~K in various magnetic fields. The field value is shown on the upper left corner of each image.  The color scale is 12~Hz. (j)-(r) Sequential MFM images of S2 (ORNL) measured at 10~K in various magnetic fields. The field value is shown on the upper left corner of each image.  The color scale is 30 Hz. Qualitatively, the same domain behavior was observed in samples from both sources. } 
    \label{fig:1}
\end{figure*}

\section{materials and methods}


Ferromagnetic \ce{MnSb2Te4} single crystals from Penn State were grown using a self-flux method \cite{Yan2019PRM, Lee2020arxiv}. 
The mixture of manganese powder (99.95\%), antimony shot (99.9999\%), and tellurium ingot (99.9999+\%) with the molar ratio of Mn:Sb:Te = 1:10:16 were loaded into alumina crucible and sealed in double evacuated quartz tubes. The mixture was heated up to 900~$^\circ$C for 12~h and then slowly cooled down to 630~$^\circ$C at a rate of 2~$^\circ$C/h and dwelt at 630~$^\circ$C for 24~h. Finally, the extra flux was removed by centrifuging at 630~$^\circ$C. The phase and crystallinity of single crystals are checked by x-ray diffraction. Single crystals from Oak Ridge National Laboratory (ORNL) were grown with similar methods \cite{Yan2019PRM}.  The ferromagnetic order with Curie temperature of 33~K was confirmed through magnetization measurements using a commercial superconducting quantum interference device (SQUID) magnetometer.

Here, we denote S1 for single crystals from Penn State, and S2 for single crystals from ORNL. Samples were cleaved in air before MFM experiments.  The MFM experiments were carried out in a homemade cryogenic magnetic force microscope using commercial piezoresistive cantilevers (spring constant $\approx 3$~N/m, resonant frequency $\approx 42$~kHz) \cite{Sass2020NL}. The homemade MFM is interfaced with a Nanonis SPM controller and a commercial phase-lock loop (SPECS). Out-of-plane magnetic field was applied via a superconducting magnet. MFM tips were prepared by depositing a nominally 100~nm Co film onto the bare tips using sputtering. The MFM signal, the change of cantilever resonant frequency, is proportional to the out-of-plane stray field gradient. Electrostatic interaction was minimized by balancing the tip-surface potential difference. Dark (bright) regions in MFM images represent attractive (repulsive) magnetization, where magnetization are parallel (antiparallel) with the tip polarization. 

\section{results and discussion}

We performed magnetic field dependent MFM measurements on S1 (6~K) and S2 (10~K), respectively. The MFM images are shown in Fig.~\ref{fig:1}. In Fig.~\ref{fig:1}(a) and Fig.~\ref{fig:1}(j), randomly shaped domains form after ZFC from 40~K. This is clearly different from the prior MFM results of antiferromagnetic \ce{MnSb2Te4} where domain walls were visualized \cite{Sass2020NL}. The typical domain size characterized by the width of the curvilinear stripes in the MFM images is approximately 1.7~$\mu$m in S1 and 2.7~$\mu$m in S2. After ZFC, we observed equal populations of up and down domains. After increasing the out-of-plane positive applied magnetic field, the domain walls propagate in such a way that the up domains (dark region) expand until they fully occupy the scanning area at around 0.1~T [Figs.~\ref{fig:1}(a)-(e) and (j)-(n)]. Consistently, S2 is magnetically saturated above 0.1 T \cite{liu2020arxiv}. Down domains nucleate as the external field is reduced from the saturation field [Figs.~\ref{fig:1}(f) and (o)] and their area increases as the field decreases [Figs.~\ref{fig:1}(f)-(i) and (o)-(r)]. The  populations of up and down domains become nearly the same again at zero applied field. These results show that both \ce{MnSb2Te4} samples are soft ferromagnets with extremely weak domain wall pinning. Moreover, the domain contrast is relatively stronger at high fields possibly due to the enhancement of the MFM tip moment. There exists  subtle difference in the domain shape between S1 and S2. Domains in S1 are more stripy than domains in S2, suggesting that the magnetism in S1 is more 2D-like. This subtle difference might be related to the different correlation length along the $c$-axis.  Further studies are needed to understand their subtle difference.

\begin{figure}[htbp]
    \centering
    \includegraphics[width=0.9\columnwidth]{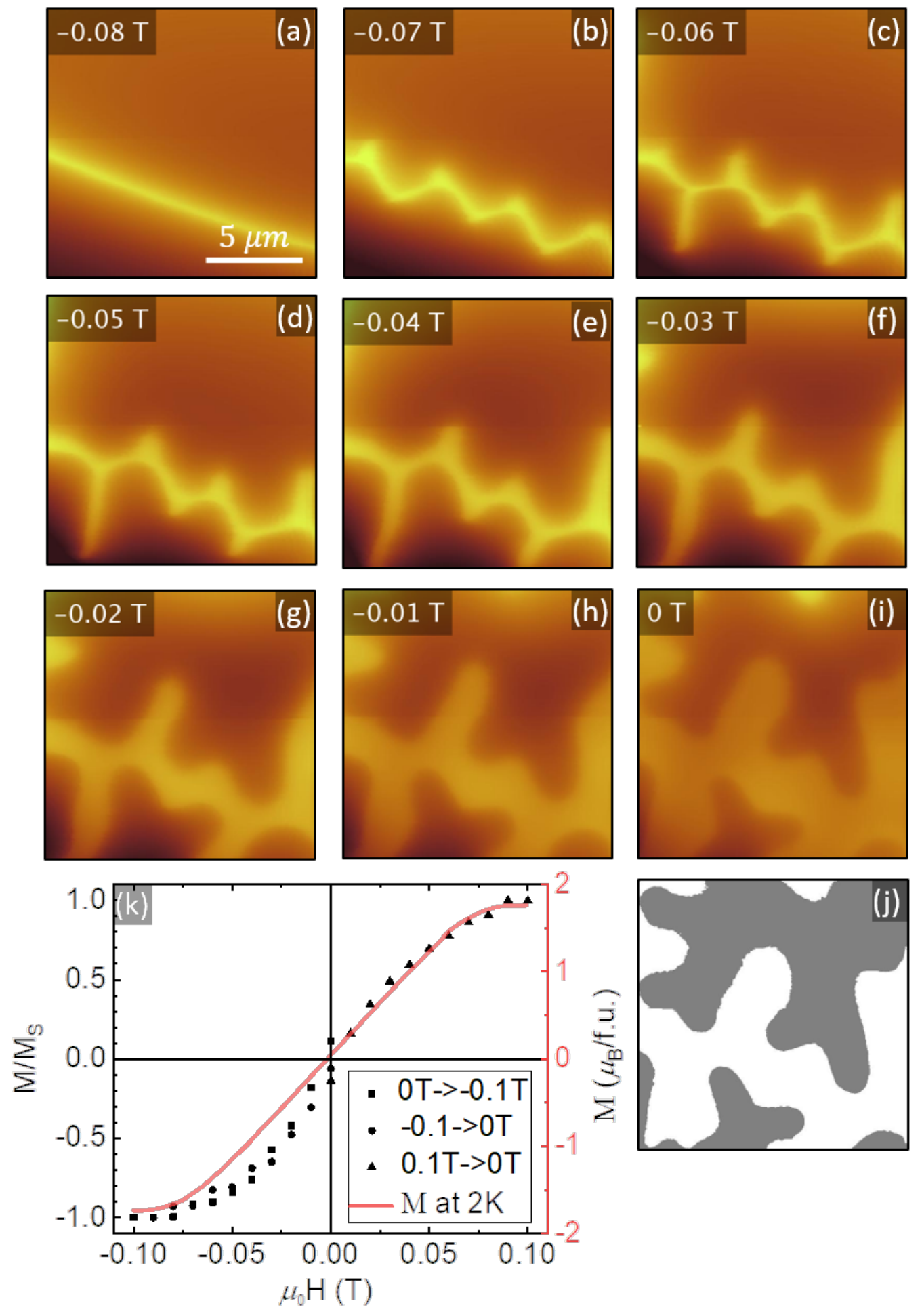}
    \caption{(Color Online) (a)-(i) Sequential MFM images of the same area of \ce{MnSb2Te4} (S2) single crystal in different magnetic fields at 5 K. The magnetic field values are labeled in the upper left corner of each image. The color scale is 20~Hz. (j) Binarized version of MFM image (i). (k) Normalized magnetization estimated from three sets of binarized field dependent MFM images. Magnetization from SQUID is plotted for comparison (red curve).}
    \label{fig:2}
\end{figure}

To demonstrate that the observed ferromagnetic domain behavior is representative of the bulk property, we performed detailed field dependent MFM measurements and domain population analysis on the S2 (ORNL) sample. Fig.~\ref{fig:2}(a)-(i) shows a negative field sweep of the S2 sample at the same location as in Fig.~\ref{fig:1}. The sample was saturated with large negative magnetic field at 5~K. The MFM tip moment was initialized in the down state. A line-shaped up domain emerges at $-0.08$~T. As the applied magnetic field further decreases, the domain starts to wiggle and branches out at the bending point in order to reduce the magnetostatic energy cost [Figs.~\ref{fig:2}(b)-(e)]. The up domain grows as the field decreases further and the area finally reaches an approximately 50\% domain population at zero-field [Figs.~\ref{fig:2}(f)-(i)]. 

\begin{figure}[htpb]
    \centering
    \includegraphics[width=0.9\columnwidth]{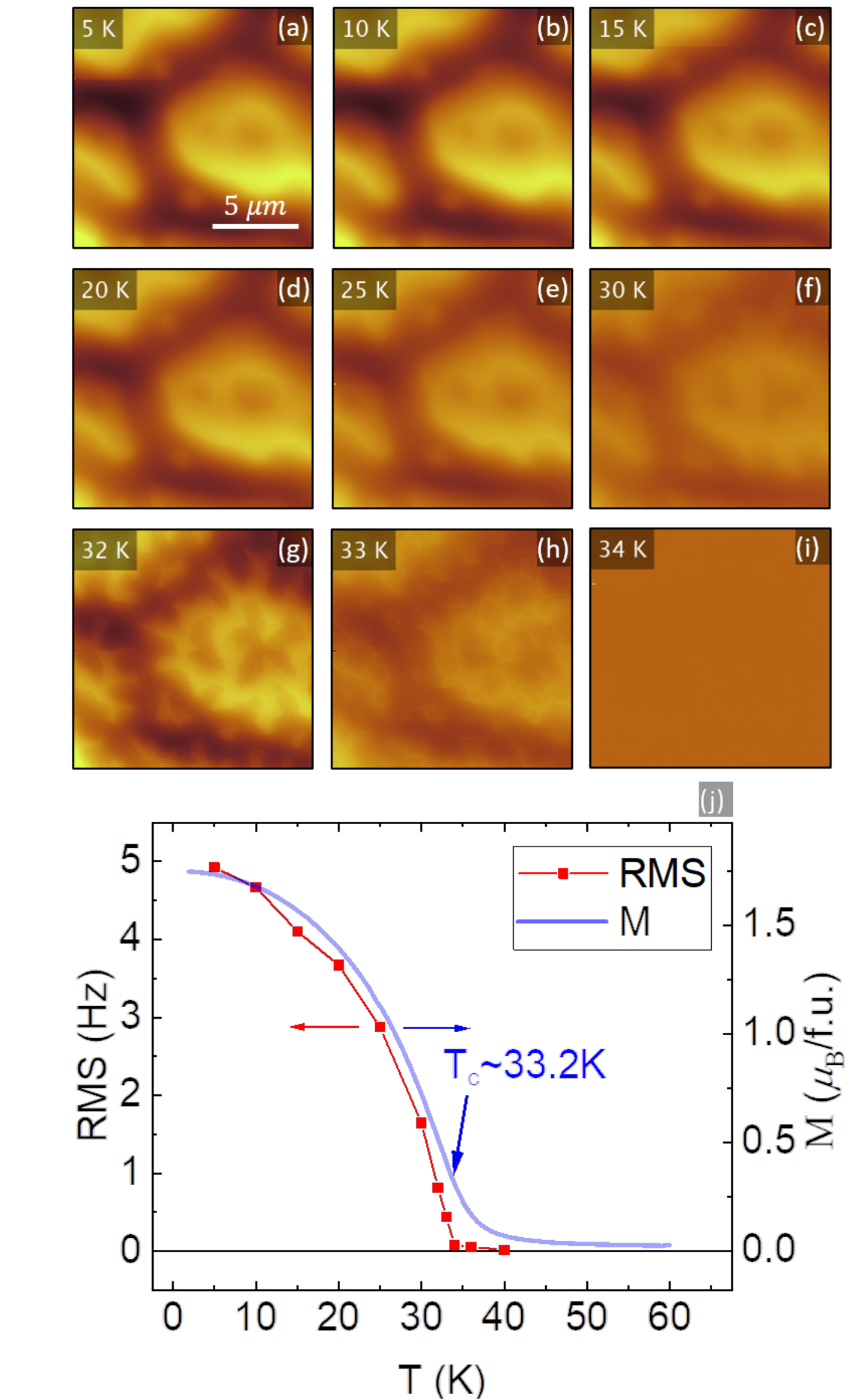}
    \caption{(Color Online) (a)-(i) Sequential zero-field MFM images of S2 at different temperatures upon warming after 0.05~T field cooling from 40~K. The temperature values are labeled in the upper left corner of each image. The scales are 20~Hz for (a)-(e) and 4~Hz for (g)-(i). (j) RMS value of MFM images as a function of temperature (red connected dots) in comparison with $M(T)$ under 0.1~T (blue curve). The Curie temperature $T_\textrm{C}$ extracted from molar susceptibility is 33.2~K as labeled in the figure.}
    \label{fig:3}
\end{figure}

Detailed examination of the domain processes reveals weak domain wall pinning in S2. To quantify the domain population and estimate the normalized magnetization, we binarize our MFM images using a threshold value approximately at the middle of the domain wall and assigning 1 ($-1$) to up (down) domains. Here, we assume that the magnetic moment is uniform everywhere inside each domain. As an example, Fig.~\ref{fig:2}(j) is the binarized version of Fig.~\ref{fig:2}(i). The normalized magnetization $M/M_\textrm{S}$ of the sample can then be estimated from the binarized MFM images from $(N_\uparrow - N_\downarrow)/(N_\uparrow + N_\downarrow)$ by counting $N_\uparrow$ and $ N_\downarrow$. Here, $M_\textrm{S}$ is the saturation magnetization and $N_\uparrow$ ($N_\downarrow$) represents the population of up (down) domains. Using the method described above, we obtain the normalized $M(H)$ data as plotted in Fig.~\ref{fig:2}(k). The external field was swept from 0~T to $-0.1$~T so that the sample was saturated, and then reduced back to 0~T. No visible hysteresis was found in the normalized magnetization $M/M_\textrm{S}$ vs.~$\mu_0H$ data, confirming the very weak domain wall pinning in \ce{MnSb2Te4}. Note that the domain patterns are different on the up-sweep and down-sweep of magnetic field, indicating random nucleation of reversed domains. Our MFM results indicate that the domain pattern is hysteretic while the total magnetization is not. 
For comparison, the $M(H)$ curve measured by SQUID is plotted in Fig.~\ref{fig:2}(k). Clearly, the estimated magnetization from local MFM imaging is in good agreement with the global magnetometry result. 

To further establish the correspondence between local (domain) and global (bulk magnetization) properties, we performed MFM measurements on sample S2 at different temperatures upon warming in zero-field after 5~mT field cooling from 40~K (above $T_\textrm{C}$). 
Bubble-shaped domains were observed, as shown in Fig.~\ref{fig:3}(a). The domain pattern remains the same below 25~K while the domain contrast decreases monotonically as the temperature increases, as shown in Figs.~\ref{fig:3}(a)-(e).
As the temperature approaches $T_\textrm{C}$, the domain contrast becomes so weak that the color scale of those images (32~K, 33~K and 34~K) was reduced to reveal the domain pattern.  
Interestingly, the domains start to deform [Fig.~\ref{fig:3}(f)] and break up into branches as shown in Fig.~\ref{fig:3}(g), indicating further reduction of anisotropy energy.  
At 33~K, just below $T_\textrm{C}$, more bubble-shaped domains form on the sample surface, which indicates a reduction of domain wall energy \cite{RudolfSchafer}. 
The domain contrast completely vanishes at and above 34~K. 
Thus, the Curie temperature is between 33~K and 34~K, in excellent agreement with $T_\textrm{C}\approx 33.2$~K estimated from the inflection point of $M(H)$ (blue curve) from SQUID measurements shown in Figs.~\ref{fig:3}(j). 

In the MFM images, the domain contrast is proportional to the stray field gradient, which scales with the saturation magnetization at each temperature \cite{Wang2018NP}. 
Since our MFM images are dominated by the domain contrast, the averaged domain contrast is proportional to the root-mean-square (rms) value of the MFM signal in each image.
Fig.~\ref{fig:3}(j) shows the temperature dependence of the rms value of MFM images Figs.~\ref{fig:3}(a)-(i). 
This curve exhibits mean field behavior which again confirms the ferromagnetic ordering of \ce{MnSb2Te4}. 
The blue curve in Fig.~\ref{fig:3}(j) shows $M(T)$ measured by SQUID with a 0.1~T out-of-plane magnetic field. 
Thus, the temperature dependence of the domain contrast at zero-field is in excellent agreement with that of the saturation magnetization, supporting the existence of long-range ordering of the ferromagnetism in \ce{MnSb2Te4}. 

\section{conclusion}
The magnetic imaging of domains in \ce{MnSb2Te4} single crystals from two different sources provides direct evidence of long-range ferromagnetic ordering in \ce{MnSb2Te4}. 
The extremely weak pinning of domain walls observed in the field dependence explains why little hysteresis was observed in bulk magnetization measurements. 
The agreement between domain population and the bulk magnetization demonstrates that the MFM measurements capture the representative domain behavior. 
Also, the temperature dependence of domain contrast agrees with the that of the saturation magnetization. 
The direct evidence of ferromagnetism in \ce{MnSb2Te4} will encourage further exploration of potential ferromagnetic Weyl physics in the \ce{MnBi2Te4}-family. 

\begin{acknowledgments}
The MFM studies at Rutgers are supported by the Office
of Basic Energy Sciences, Division of Materials Sciences and Engineering, U.S. Department of Energy under Award No.~DE-SC0018153.  Work at ORNL was supported by the U.S. Department of Energy, Office of Science, Basic Energy Sciences, Materials Sciences and Engineering Division.
Support for crystal growth and characterization at Penn State was provided by the National Science Foundation through the Penn State 2D Crystal Consortium-Materials Innovation Platform (2DCC-MIP) under NSF Cooperative Agreement DMR-1539916.
\end{acknowledgments}

\bibliography{MST}

\end{document}